# Crystallizing Kagome artificial spin ice

Wen-Cheng Yue[1,2,+], Zixiong Yuan[1,2,+], Yang-Yang Lyu[1], Sining Dong[1,*], Jian Zhou[2], Zhi-Li Xiao[3,4,*], Liang He[2], Xuecou Tu[1,5], Ying Dong[6], Huabing Wang[1,5,*], Weiwei Xu[1], Lin Kang[1,5], Peiheng Wu[1,5], Cristiano Nisoli[7], Wai-Kwong Kwok[3], and Yong-Lei Wang[1,2,5,*]

[1]*Research Institute of Superconductor Electronics, School of Electronic Science and Engineering, Nanjing University, Nanjing, 210023, China*

[2]*Jiangsu Provincial Key Laboratory of Advanced Photonic and Electronic Materials, School of Electronic Science and Engineering, Nanjing University, Nanjing 210093, China*

[3]*Materials Science Division, Argonne National Laboratory, Argonne, Illinois 60439, USA*

[4]*Department of Physics, Northern Illinois University, DeKalb, Illinois 60115, USA*

[5]*Purple Mountain Laboratories, Nanjing, 211111, China*

[6]*Research Center for Quantum Sensing, Zhejiang Lab, Hangzhou, Zhejiang, 311121, China*

[7]*Theoretical Division and Center for Nonlinear Studies, Los Alamos National Laboratory, Los Alamos, New Mexico 87545, USA*

+ Authors contribute equally

* Correspondence to: sndong@nju.edu.cn; xiao@anl.gov; hbwang@nju.edu.cn; yongleiwang@nju.edu.cn

**Artificial spin ices are engineered arrays of dipolarly coupled nanobar magnets. They enable direct investigations of fascinating collective phenomena from their diverse microstates. However, experimental access to ground states in the geometrically frustrated systems has proven difficult, limiting studies and applications of novel properties and functionalities from the low energy states. Here, we introduce a convenient approach to control the competing diploar interactions between the neighboring nanomagnets, allowing us to tailor the vertex degeneracy of the ground states. We achieve this by tuning the length of selected nanobar magnets in the spin ice lattice. We demonstrate the effectiveness of our**

**method by realizing multiple low energy microstates in a Kagome artificial spin ice, particularly the hardly accessible long range ordered ground state – the spin crystal state. Our strategy can be directly applied to other artificial spin systems to achieve exotic phases and explore new emergent collective behaviors.**

Artificial spin ices (ASIs) are exemplar material-by-design systems with intriguing physical phenomena such as geometrical frustration [1–6], monopole-like excitations [7–10], Coulomb phase [10,11] and phase transitions [12–14]. They lead to novel functionalities with great potential for applications, such as low-power data storage [15], encryption devices [16] and advanced computations [17–19]. As one of the simplest ASI structures, the Kagome ASI has attracted extensive attention [2,7–9,12,13,20–29], because it is highly frustrated and contains a rich phase diagram. Theoretical investigations suggest four thermal phases with reducing temperatures [13,21,28,30]: a high temperature paramagnetic state (PM phase); a spin liquid with correlated spins satisfying the Kagome ice rule ('two in/one out' or 'two out/one in') but with neither charge nor spin ordering (SL1 phase); a long-range ordered charge crystal in a disordered spin liquid state (SL2 phase); as well as a spin crystal state in which the spins have long-range ordering (LRO phase) as the lowest temperature state. These thermal phases were used to understand the temperature dependent magnetotransport results in honeycomb structures of nanowire networks (which is sometimes also called Kagome ASI, because the moments of the nanowires satisfy the Kagome ice rule), where the LRO spin crystal state may contribute to the topological Hall signals at the lowest temperatures [30]. However, previous investigations have shown that direct experimental access of the spin crystal states of a Kagome ASI is challenging [12,31]. With the exception of the magnetic writing approach [31], no LRO spin crystal state has been unambiguously visualized in a Kagome ASI consisting of fully disconnected nanobar magnets, hindering investigations of emergent phenomena and phase transitions from its low energy manifolds.

The collective properties of ASIs are directly associated with their lattice geometries, and are intimately governed by the competing dipolar interactions between their constituent elements, i.e., the single-domain nanobar magnets [16,32–34]. The difficulty of obtaining the LRO state of a Kagome ASI originates from the extensive degeneracy of the ground state, resulting from the high frustration of the tri-leg vertices. The ground state of a Kagome ASI exhibits neither charge nor spin order when only nearest-neighbor interactions are considered [21,35]. Further-neighbor interactions induce charge and/or spin ordering [13,21,28]. However, these longer-distance interactions are much weaker than that from the nearest-neighbor. As a result, the properties of artificial spin ices are mostly dominated by the nearest-neighbor interactions. Recently, utilizing the micromagnetic nature of connected vertices in the honeycomb structure of nanowire networks, the LRO spin crystal phase of Kagome ices was realized for the first time, in which notches were introduced to reduce the vertex degeneracy [36]. More recently, asymmetric bridges were introduced to break the six-fold symmetry of Kagome ASI's vertices, leading to a direct real-space imaging of the phase transitions [37]. However, these strategies are not applicable to Kagome ASIs consisting of disconnected nanobar magnets. In this letter, we develop a new method to tailor the vertex degeneracy and the ground states of a Kagome ASI of disconnected magnetic nanobars, and we directly present the phase transition from the SL1 liquid to LRO crystal state. Unlike the recently realized connected Kagome ASI structure [36,37], in which the coupling is dominated by short range exchange interaction, our disconnected ASIs maintain their long range dipolar interaction, therefore, allowing us to directly evaluate the critical role of the nearest and/or next nearest neighbor interactions, which is crucial for understanding the phase transitions between the various low energy manifolds of ASIs.

In square ASIs, the six vertex configurations satisfying the spin ice-rule are divided into two types according to energy [1,2,10,11,14,17,18,22,24,29,38–44]. The lowest energy configuration is two-fold degenerate, leading to the formation of a long-range ordered ground state [1,38–41]. The low degeneracy of 2 in a square ASI is induced by the non-equal interactions between the nanobar magnets at each vertex. Following this notion, we developed a method to achieve the LRO spin ice state of a regular Kagome ice by inducing non-equal interactions between the three nanobar magnets at a vertex, which reduces the ground state degeneracy of a Kagome vertex from 6 to 2. As shown in Fig. 1(a), we increase the length of one of the three nanobar magnets ($\alpha$) in each vertex, while maintaining the length of the other two magnets ($\beta$) and the lattice constant (see Fig. S1 of Supplemental Material for the detailed arrangement of $\alpha$ and $\beta$ magnets in the lattice [45]). This breaks the three-fold rotation symmetry of the vertex. The interactions between the three nanobar magnets at each vertex are no longer equivalent. As shown in Fig. 1(b), the original six-fold degenerate vertices are divided into two groups of Type K-I and K-II configurations with different energies. We denote the interaction energy of the frustrated magnet pair between two $\beta$ nanomagnets in Type K-I vertices as $J_1$, while that between $\alpha$ and $\beta$ nanomagnets in Type K-II vertices as $J_2$ [Fig. 1(b)]. Since each vertex satisfying the Kagome ice rule contains only one frustrated magnet pair, $J_1$ and $J_2$ also represent the energies of the Type K-I and K-II vertices, respectively. Because the endpoints of the lengthened α nanomagnet are closer to the vertex center, $J_1$ is lower than $J_2$, resulting in the two-fold degenerate Type K-I vertices to be the ground state configuration. As shown in Fig. 1(a), when all the vertices are satisfied to be in the Type K-I ground state configuration, a LRO spin crystal emerges. We can further tailor the energy difference between Type K-I and K-II vertices by varying the length ($L_α$) of the α nanomagnet. Figure 1(c) presents the vertex energy evolution of Type K-I and K-II vertices as a function of $L_α$, calculated from micromagnetic simulation using Mumax3 [46]. It shows that $J_2$ increases with $L_α$ while $J_1$ remains constant. Thus, the energy difference $J_2$- $J_1$ between Type K-I and K-II vertices increases with $L_α$. Therefore, varying the length of the α nanomagnets allows us to regulate the effective temperature, similar to the effect of tuning the vertex notch in the connected honeycomb structures [36]. This enables us to tailor the phases in a fully disconnected Kagome ASI.

To experimentally validate this approach, we fabricated Kagome ASIs with Permalloy nanomagnets and with a series of $L_α$ values (220 nm, 270 nm, 320 nm, 370 nm, 420 nm and 440 nm) for the α nanomagnets (see SEM images in Fig. S2 of Supplemental Material [45]). The length ($L_β$) of the β magnets is fixed at 220 nm, and the lattice constant is $a$ = 640 nm for all samples [Fig. 1(a)]. The width and thickness of all the nanomagnets are 80 nm and 15 nm, respectively. Details of the sample fabrication process and parameters can be found in Supplemental Material [45]. A demagnetization procedure (see Methods) lasting 72 hours was performed to obtain the low-energy states [11,47]. Figures 1(d) and 1(e) show the SEM images of the samples with $L_α$ = 220 nm and 420 nm, respectively. The corresponding magnetic force microscopy (MFM) images are displayed in Figs. 1(f) and 1(g), respectively, which allow us to determine the magnetic moment (or spin) configurations [see arrows in Figs. 1(f) and 1(g)]. The results show that all the vertices in all measured samples satisfy the Kagome ice rule ('two in/one out' or 'two out/one in'). This indicates our demagnetization procedure successfully brought the samples into the low energy ice rule manifold. The conventional Kagome ASI with $L_α = L_β$ [Fig. 1(d)] exhibits disordered spin and charge configurations [Fig. 1(f)], consistent with the frozen spin liquid SL1 state. In contrast, the modified Kagome ASI with $L_α > L_β$ (Fig. 2b) exhibits perfect spin and charge ordering [Fig. 1(g)], leading to a successful realization of the LRO spin crystal ground state.

The transition from SL1 phase to LRO spin crystal phase is demonstrated by the MFM images of samples with increasing $L_α$ values in Figs 2(a)-2(d) (results for all six measured samples are shown in Fig. S3 of Supplemental Material [45]). The corresponding maps of vertex distributions in Figs. 2(e)-2(h) shows that domains of crystallization become larger with $L_α$. For the sample with $L_α = 220$ nm (the conventional Kagome ASI), the Type K-I and K-II vertices are degenerate, leading to a disordered magnetic state with vertex populations of 33.75% and 66.25% for Type K-I and K-II vertices, respectively. This is consistent with the configurational (or random) populations of 1/3 and 2/3 for Type K-I and K-II vertices, as expected for a spin liquid, and proves that our demagnetization procedure effectively brought the system into an effective thermal equilibrium state. When $L_α > 200$ nm, the Type K-II vertices become an excited state. With increasing $L_α$, the energy difference between Type K-I and Type K-II vertices increases [Fig. 1(c)] and thus the population of Type K-I vertices gradually increases [Fig. 3(a)]. As shown in Figs. 2(e)-2(h), ordered domains of Type K-I vertices emerge and grow with increasing $L_α$. For the sample with large $L_α$, such as 420 nm [Fig. 2(h)], 92.6% of the vertices are in the Type K-I ground state, and domain walls comprised of excited Type K-II vertices (red) are clearly visible. Figure 3(b) displays the normalized vertex populations as a function of the excitation energy $J_2$-$J_1$ of Type K-II vertices. It shows that the population of the ground state Type K-I vertices rises rapidly at small $J_2$-$J_1$ values, and saturates when the values of $J_2$-$J_1$ become large. The statistical results from Monte Carlo simulations [Figs. 3(c) and 3(d)], with a thermal annealing protocol and considering only nearest-neighbor interactions [see Supplemental Material [45]], match nicely with those from experiments [Figs. 3(a) and 3(b)]. This suggests that the transition from SL1 phase to LRO spin crystal state is dominated by the nearest-neighbor interactions.

We can further elucidate the spin and charge ordering using magnetic structure factors [47]. Maps of the magnetic spin structure factor are shown in Figs. 2(i)-2(l) (results for all measured samples can be found in Fig. S3 of Supplemental Material [45]). For the conventional Kagome ASI ($L_\alpha$ = 220 nm), the magnetic spin structure factor map shows structured diffusive pattern [Fig. 2(i)], consistent with previous results of SL1 phase [13,36]. With gradually increasing $L_\alpha$, we can clearly observe the emergence and enhancement of Bragg peaks [Figs. 2(k) and 2(l)]. This demonstrates again the transition from a spin liquid state into a long-range ordered spin crystal. These Bragg peaks show split-peak structures whose number reduces with increasing $L_\alpha$, as shown in the insets of Figs. 2(k) and 2(l). These split Bragg peaks originate from scattering between domains, as demonstrated by the structure factor maps of the artificial domain configurations in Fig. S5 of Supplemental Material [45]. The number of domains decreases when they merge, resulting in reduced number of split Bragg peaks. This suggests that the texture of Bragg peaks could be used as a qualitative parameter to investigate ordering and domain formation in ASIs, e.g., we could estimate the relative sizes of domains from the number of split Bragg peaks.

Previous investigations of conventional Kagome ASIs revealed local ordering of magnetic charges [22–24]. Figs. 4(a)-4(d) are maps of magnetic charge configurations corresponding to the spin and/or vertex configurations in Figs. 2(e)-3(h) (see Fig. S6 of Supplemental Material [45] for more data). We can see that the charge domains of the two-fold degenerate phases (green and yellow) grow with increasing $L_\alpha$. When comparing the charge distributions [Figs. 4(a)-4(d)] to the vertex or spin distributions [Figs. 2(e)-2(h)], we can see that their profiles match very well with each other for the samples with large $L_\alpha$ values [Figs. 2(h) and 4(d)], i.e., the charge ordering is embedded in the spin ordering for spin crystal state. However, the charge and vertex distributions deviate as the number of Type K-II vertices increases [Figs. 2(e) and 4(a)]. This is because the charge configurations are not tied to the types of vertex configurations in the spin liquid state. A similar behavior is observed when we compare the magnetic spin structure factor maps [Figs. 2(i)-2(l)] with the magnetic charge structure factor maps [Figs. 4(e)-4(h)]. For large $L_\alpha$, both spin [Fig. 2(l)] and charge [Fig. 4(h)] structure factor maps exhibit clear Bragg peaks with the same peak splitting [insets of Fig. 2(l) and Fig. 4(h)]. However, for small $L_\alpha$ values, e.g., for the conventional Kagome ASI with $L_\alpha = L_\beta = 220$ nm, the spin structure factor map is structured but completely diffusive [Fig. 2(i)], while in contrast, the charge structure factor maps display clear Bragg peaks, although they are broad [Fig. 4(e)]. This indicates the emergence of charge ordering in the spin liquid state. As mentioned earlier, charge ordering should not appear when there are only nearest-neighbor interactions and the six-fold symmetry is not broken [21,35]. Our Monte Carlo simulations with only nearest-neighbor interactions show that there are no Bragg peaks in the magnetic charge structure factor map for conventional Kagome ASI (Fig. S7g of Supplemental

Material [45]). This suggests that further-neighbor interactions beyond nearest-neighbors play a role in the liquid state. On the other hand, the consistency between experiment [Fig. 4(h)] and simulation (Fig. S7k of Supplemental Material [45]) for samples with large $L_a$ unambiguously suggests the LRO spin crystal phase can be established with only nearest-neighbor interactions.

We have shown that the local interactions of an ASI have significant impact on their collective behavior and ultimately affect the properties of the entire system. The length of selective nanobar magnets can be used as a convenient knob to tune the local coupling strength, which enables us to access various low-energy manifolds and phase transitions in a fully disconnected ASI. We proved that although the next-nearest neighbor interaction plays a notable role through the magnetic charge structure factor maps, the crystallization of the Kagome ice structure with reduced vertex degeneracies is dominated only by the nearest neighbor interaction. This method could be used to manipulate the frustration in ASIs for attaining even more exotic ground state phases (see Figs. S8 of Supplemental Material [45]). It would also allow us to realize new magnetic configurations that are hard to access with magnetization method, e.g., to design novel ASIs with composite ground states in which different low energy states co-exist in the same sample, as illustrated by the hybrid ground state of both liquid and crystal in Fig. S9 of Supplemental Material [45]. This would allow us to investigate phase transitions between these new types of low energy states. Furthermore, kinetics becomes topologically protected in the SL2 phase and in the LRO state [48], and our structural modifications can be employed to fine-tune different kinetic regimes in thermal realizations or under field inversions. In addition, this method is also applicable to other types of artificial spin ices, leading to new opportunities to explore more exotic collective phenomena, such as novel phases and phase transitions. It could also be combined with the other structure modification strategies, such as lattice transformation [49,50], to design new ASIs with controllable degeneracies. This method, maintaining the dipolar coupling between disconnected nanomagnets, could result in different spin dynamic properties and magnonic applications [51,52] than those in connected systems [36,37]. Moreover, our approach avoids complex analysis of the micromagnetic structures such as the domains/domain-walls in the vertices of connected systems, and thus offers a simpler Kagome model which can connect with a variety of systems outside of magnetism, such as metal organic frameworks [53] and mechanical metamaterials [54, 55]. Last but not least, the observed Bragg peaks' splitting induced by spin scattering between domains could lead to an alternative way to investigate domain/domain wall formation in ASIs.

**Acknowledgment**

This work is supported by the National Key R&D Program of China (2018YFA0209002 and 2021YFA0718802), the National Natural Science Foundation of China (61771235, 61971464, 61727805, 11961141002, 62101243), Jiangsu Excellent Young Scholar program (BK20200008) and Jiangsu Shuangchuang program. Part of the project design (Z.L.X.) and manuscript writing (Z.L.X. and W.K.K.) are supported by the U.S. Department of Energy, Office of Science, Basic Energy Sciences, Materials Sciences and Engineering. Z.L.X. also acknowledges support from the National Science Foundation under Grant No. DMR-1901843 for his efforts on data analysis. Y.D. acknowledges support from the Major Scientific Research Project of Zhejiang Lab (2019MB0AD01), the Center initiated Research Project of Zhejiang Lab (2021MB0AL01) and the Major Project of Natural Science Foundation of Zhejiang Province (LD22F050002).

**Figure captions**

**Fig.1**. Tunable Kagome artificial spin ice. (a) Design of tunable Kagome artificial spin ice with extended length $L_\alpha$ of $\alpha$ magnets and fixed length $L_\beta$ of $\beta$ nanomagnets. (b) Six low energy vertex configurations satisfying the Kagome ice rule are separated into two groups based on energies. $J_1$ and $J_2$ are the coupling strengths of frustrated magnet pairs. (c) Evolution of the vertex energies as a function of $L_\alpha$. (d) and (e) SEM images of Kagome artificial spin ices with $L_\alpha = 220$ nm (d) and $L_\alpha = 420$ nm (e), respectively. Scale bar, 500 nm. (f) and (g) MFM images corresponding to (d) and (e), respectively. Arrows indicate spin configurations. Scale bar, 500 nm.

**Fig.2**. Transition from liquid to crystal. (a)-(d) MFM images for samples with $L_\alpha = 220$ nm, 270 nm, 320 nm and 420 nm, respectively. Scale bar, 2 μm. (e)-(h) spin configurations and vertex distributions extracted from (a)-(d). Type K-I and K-II vertices are shown in blue and red, respectively. (i-l) corresponding maps of magnetic structure factors calculated respectively from the spin configurations in (e)-(f). Insets of (k) and (l) show expanded views of the split Bragg peaks.

**Fig.3**. Vertex populations. (a) Statistics of the vertex populations from experiments. (b) Normalized vertex populations by the configurational populations of 1/3 (Type K-I) and 2/3 (Type K-II) as a function of the excitation energy of Type K-II vertices. The top labels show the $L_\alpha$ values of the corresponding data points. The dashed blue horizontal line at 1 represents the random populations, above and below which indicate favorable and unfavorable occupations, respectively. (c) and (d) the corresponding statistical results from Monte Carlo simulations. The data were extracted from Fig. S4.

**Fig.4**. Magnetic charge ordering. (a)-(d) magnetic charge distributions corresponding to MFM images in Figs. 3(a)-3(d) respectively. Green and yellow denote two phases of magnetic charge ordering. Scale bar, 2 μm. (e)-(h) maps of magnetic charge structure factors associated with (a-d). Insets show expanded views of the split Bragg peaks.

**Figure 1**

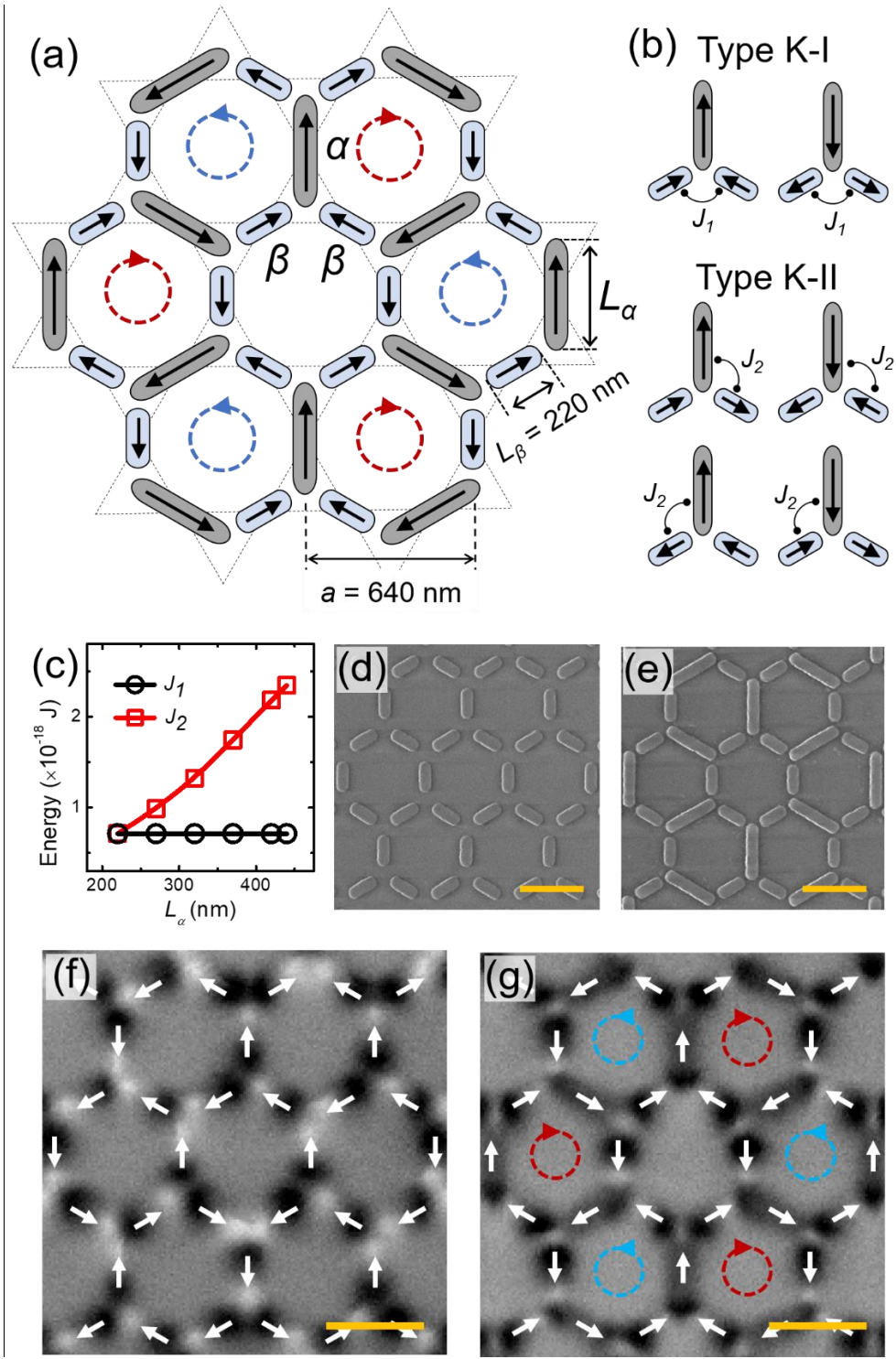

**Figure 2**

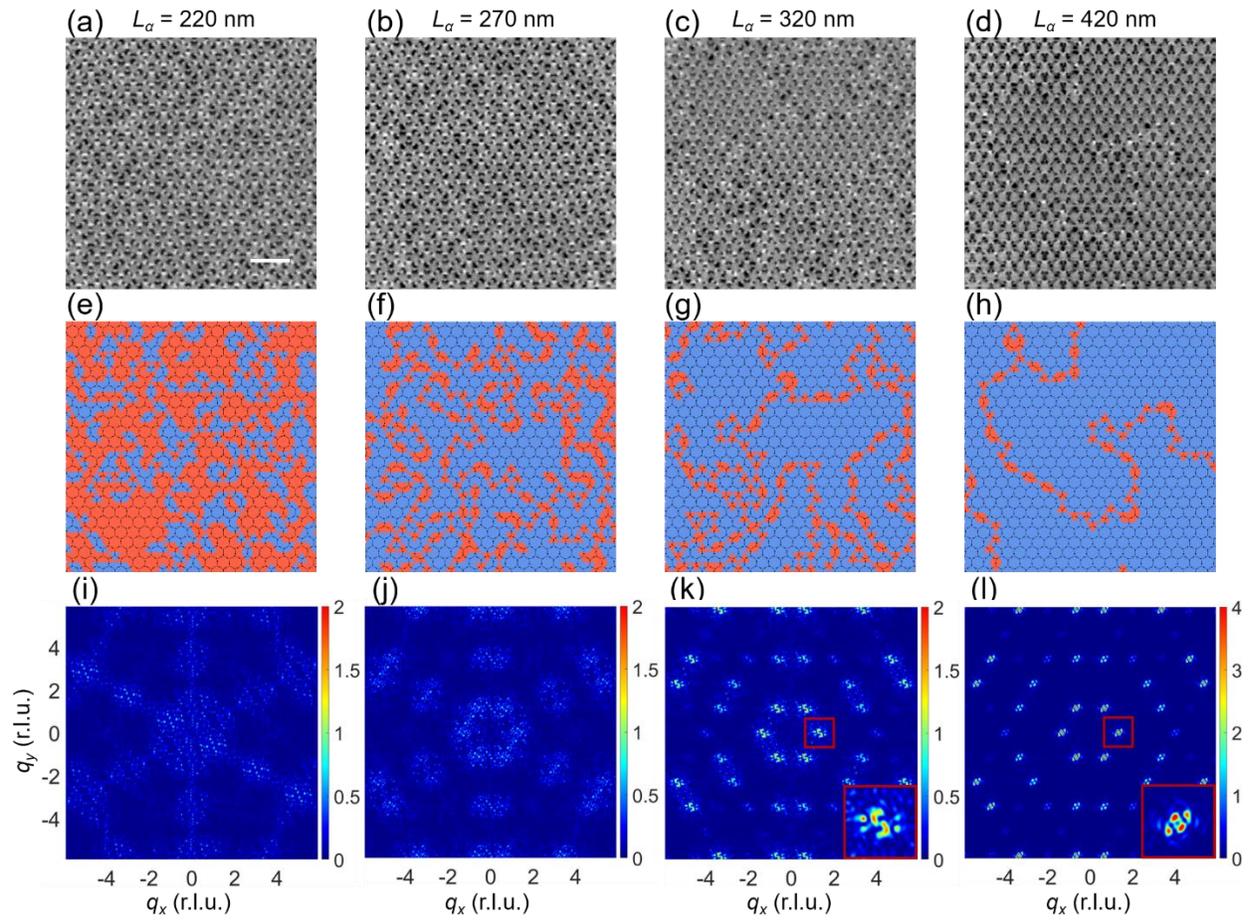

**Figure 3**

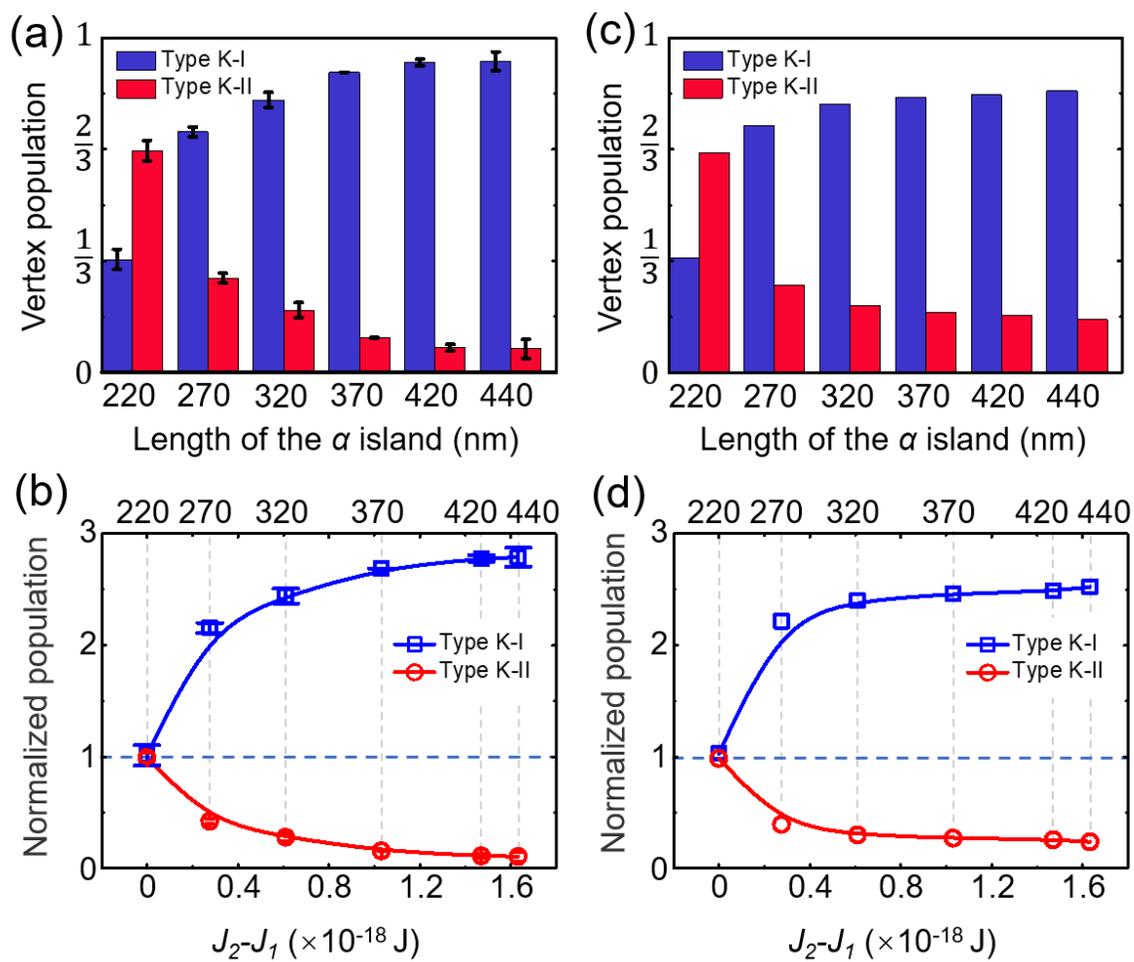

**Figure 4**

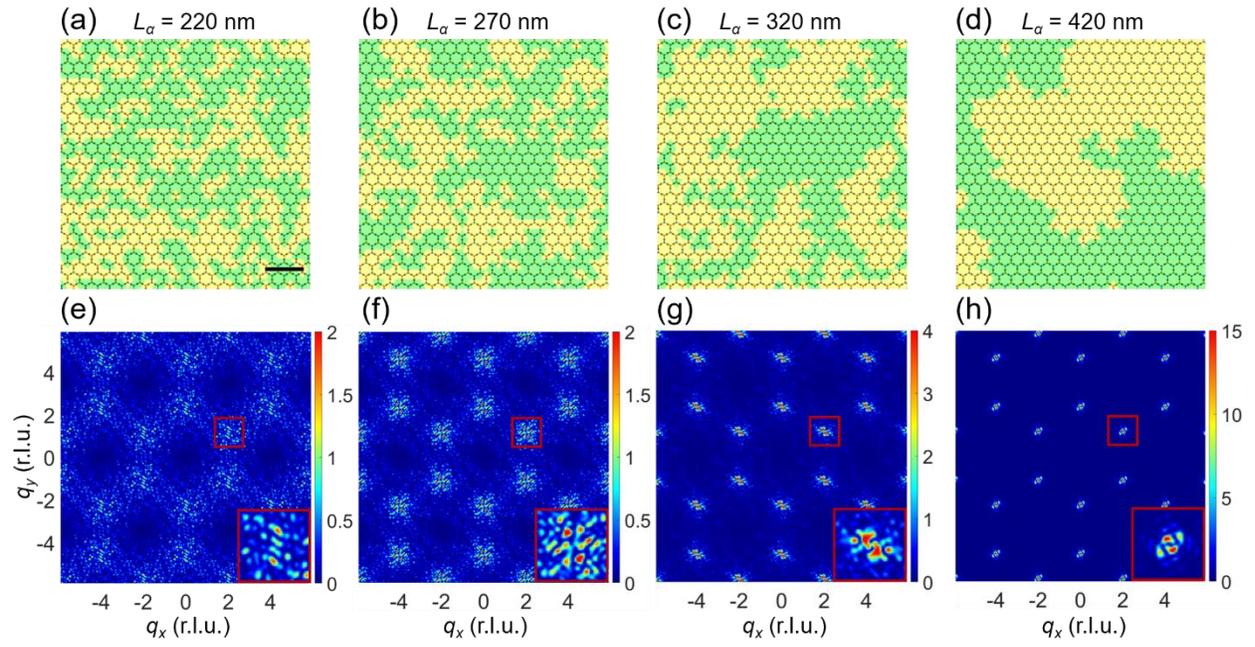

# Supplemental Materials

## Crystallizing Kagome artificial spin ice

Wen-Cheng Yue[1,2,+], Zixiong Yuan[1,2,+], Yang-Yang Lyu[1], Sining Dong[1,*], Jian Zhou[2], Zhi-Li Xiao[3,4,*], Liang He[2], Xuecou Tu[1,5], Ying Dong[6], Huabing Wang[1,5,*], Weiwei Xu[1], Lin Kang[1,5], Peiheng Wu[1,5], Cristiano Nisoli[7], Wai-Kwong Kwok[3], and Yong-Lei Wang[1,2,5,*]

[1]Research Institute of Superconductor Electronics, School of Electronic Science and Engineering, Nanjing University, Nanjing, 210023, China

[2]Jiangsu Provincial Key Laboratory of Advanced Photonic and Electronic Materials, School of Electronic Science and Engineering, Nanjing University, Nanjing 210093, China

[3]Materials Science Division, Argonne National Laboratory, Argonne, Illinois 60439, USA

[4]Department of Physics, Northern Illinois University, DeKalb, Illinois 60115, USA

[5]Purple Mountain Laboratories, Nanjing, 211111, China

[6]Research Center for Quantum Sensing, Zhejiang Lab, Hangzhou, Zhejiang, 311121, China

[7]Theoretical Division and Center for Nonlinear Studies, Los Alamos National Laboratory, Los Alamos, New Mexico 87545, USA

+ Authors contribute equally

* Correspondence to: sndong@nju.edu.cn; xiao@anl.gov; hbwang@nju.edu.cn; yongleiwang@nju.edu.cn

**Sample fabrication.**

We fabricated the nanomagnet arrays on silicon substrates with a silicon nitride layer of 200 nm. A bilayer electron-beam (e-beam) resist of PMMA 495 (100 nm) and PMMA 950 (80 nm) was coated onto the substrate. The nanomagnet arrays were then patterned using e-beam lithography, followed by e-beam evaporation of 15 nm thick permalloy ($Ni_{0.8}Fe_{0.2}$) at a deposition rate of 0.3Å/s. A 3 nm thick aluminum capping layer was deposited on the top to prevent oxidation of the permalloy. Each array is 100 μm by 100 μm in size and contains approximately ~$8\times10^4$ nanomagnets.

**Demagnetization protocol.**

The sample was mounted on a motor rotating at 2000 RMP. An oscillating in-plane magnetic field (sine wave with period of 40 s) was applied to the sample. The amplitude of the oscillating magnetic field decreased slowly from 1000 Oe (well above the saturation field of our nanomagnets) to zero field in 72 hours.

**Micromagnetic simulations.**

The micromagnetic simulations were performed using Mumax3 [46], with the following material parameters for permalloy: the exchange constant of $1.3\times10^{-11}$ J/m, the saturation magnetization of $8.6 \times 10^5$ A/m, and the Gilbert damping of 0.01. The mesh size is $2 \times 2 \times 2$ $nm^3$. We consider a given pair of neighboring nanobar magnets. The ground state configuration is 'one-in, one out' or 'head to tail' configuration, and the frustrated high energy configuration is 'both in' or 'both out' configuration. We extract the energies of the ground state configuration and the frustrated high energy configuration from micromagnetic simulations. The interaction energy (J1 or J2) of the magnet pair is obtained by subtracting the frustrated high energy from the ground state energy. In this case, the ground state energy of the entire pair of two neighboring magnets is zero.

**Monte Carlo simulated thermal annealing to the ground state.**

Our Monte Carlo simulations were performed by using a single spin flip algorithm on 30×30×3 Kagome lattice sites with free boundary conditions. In the simulation, the Hamiltonian is defined as $H = -\sum_{\langle i,j \rangle} J_{ij} S_i S_j$, where $S_i$ and $S_j$ are Ising variables on the sites $i$ and $j$, and $J_{ij}$ is the coupling strength between the two sites. Here, we only consider the interaction between the nearest neighbors. There are two values $J_1$ and $J_2$ for $J_{ij}$, which are given by micromagnetic simulations in Fig. 1b. We performed a thermal annealing protocol to obtain the ground states. The system is cooled from $T = 2$ to 0 in $2.5 \times 10^5$ Monte Carlo steps for all the samples. This process mimics a thermal annealing process under the same (finite) annealing time (the same simulation steps) for all the samples. In this case, the sample with larger $J_2$-$J_1$ relaxes to ground state faster, in other words, under the same annealing time (or the same MC steps), the sample with larger $J_2$-$J_1$ is closer to a pure Type K-I ground state.

**Magnetic spin structure factor**.

The magnetic spin structure factor was calculated by following reference [47]. We define a perpendicular spin component $S^{\perp}$ as:

$$\vec{S}^\perp = \vec{S} - (\hat{q} \cdot \vec{S})\, \hat{q}$$

where $\hat{q}$ is the unit scattering vector: $\hat{q} = \dfrac{\vec{q}}{\|\vec{q}\|}$

So for every vector $q = (q_x, q_y)$, the intensity $I(q)$ scattered at location $q$ in reciprocal space is written as:

$$I(\vec{q}) = \frac{1}{N} \sum_{(i,j=1)}^{N} \vec{S}_i^\perp \cdot \vec{S}_j^\perp \exp(i\vec{q} \cdot (\vec{r}_i - \vec{r}_j))$$

$I$ is the magnetic spin structure factor that we calculate for the interval $(q_x, q_y) = [-5.9\pi, -5.9\pi]-[5.9\pi, 5.9\pi]$ in 513×513 steps.

**Magnetic charge structure factor.**

The magnetic charge structure factor is calculated in a similar way as the spin structure factor. We define the direction of magnetic charge to be in the **Z** direction, perpendicular to the sample plane. For each vertex, we consider the total charge $\vec{C}$ at the vertex center. For every vector $q = (q_x, q_y)$ the intensity $I(q)$ is given by

$$I(\vec{q}) = \frac{1}{N} \sum_{(i,j=1)}^{N} \vec{C}_i \cdot \vec{C}_j \exp(i\vec{q} \cdot (\vec{r}_i - \vec{r}_j))$$

where $N$ stands for the number of charges.

In order to simplify the calculation, the equation is divided into two parts, each containing site i and j respectively, and is rewritten as

$$I(\vec{q}) = \frac{1}{N} \left( \sum_{i=1}^{N} \vec{C}_i \exp(i\vec{q}\vec{r}_i) \right) \cdot \left( \sum_{j=1}^{N} \vec{C}_j \exp(-i\vec{q}\vec{r}_j) \right)$$

The above equation can also be written as

$$I(\vec{q}) = \frac{1}{N}(\sum_{i=1}^{N}\vec{C}_i \cos(\vec{q}\cdot\vec{r}_i) + i\sum_{i=1}^{N}\vec{C}_i \sin(\vec{q}\cdot\vec{r}_i)) \cdot (\sum_{j=1}^{N}\vec{C}_j \cos(\vec{q}\cdot\vec{r}_j) - i\sum_{j=1}^{N}\vec{C}_j \sin(\vec{q}\cdot\vec{r}_j))$$

Then we obtain the equation

$$I(\vec{q}) = \frac{1}{N}(\vec{A}+i\vec{B})\cdot(\vec{A}-i\vec{B}) = \frac{1}{N}(\vec{A}^2+\vec{B}^2)$$

where $\vec{A} = \sum_{i=1}^{N}\vec{C}_i \cos(\vec{q}\cdot\vec{r}_i)$ and $\vec{B} = \sum_{i=1}^{N}\vec{C}_i \sin(\vec{q}\cdot\vec{r}_i)$. $I$ is now the quantity of magnetic charge structure factor that we calculate for the interval $(q_x, q_y) = [-5.9\pi, -5.9\pi] - [5.9\pi, 5.9\pi]$ in 513×513 steps.

**Comparisons to other recent bicomponent/hybrid ASIs.**

Recently, bicomponent square ASIs constructed with nanobar magnets of either different materials [44] or different shapes [25,43] were introduced for reconfigurable magnonic applications. Although we also utilize bicomponent elements (nanobar magnets with different lengths) in the Kagome ASI, our tactic differs fundamentally from those reported. There, the ground state configurations and multiple metastable configurations of square ASIs are realized through well-designed magnetization processes, which utilize the distinct magnetization properties, such as different magnetization reversal fields of different types of nanobar magnets [25,43,44]. The interactions between nanomagnets are not vital in those processes, therefore tuning the ASIs' configurations using polarizing magnetic fields does not directly reveal their collective behavior. In contrast, the interaction between nanomagnets in our modified Kagome ASIs is crucial for achieving their collective low energy manifolds. Varying the length and the arrangements of the selected nanobar magnets allows us to tailor their interaction and/or coupling strength, enabling us to obtain distinct ground states, e.g., the two exotic ground states in Fig. S8 and a hybrid ground state with both liquid and crystal states in Fig. S9.

In addition, recently site-specific exchange-bias fields were introduced to tune the ground state in a hybrid square ASI by introducing local pinning of nanomagnets through exchange bias effect [42]. In that work the magnetization state of selected nanomagnets were fixed or pinned by exchange bias. These fixed nanomagnets behave like controllable defects, therefore creating a new energy landscape for the rest of the nanomagnets. The ground state of the square ASI was tuned by introducing a designed pinning landscape. In our work, we modify the length of selected nanomagnets, which does not pin or fix any magnets. That is, we tailor the ground state configuration by tuning the interactions between the nanomagnets and not by producing a fixed new energy pinning landscape. On the other hand, as mentioned in Ref [42], high temperatures can destroy the interfacial exchange bias, and therefore the method of introducing pinning with exchange bias cannot be used to investigate the thermal properties, e.g., the sample cannot go through thermal annealing. Although in our experiments we realize the ground state using a demagnetization procedure, our sample in principle can be used to explore the thermal properties at high temperatures.

**Comparisons to rectangular ASIs.**

Previous works in Refs [49,50] modified the length of the lattice constant of a square ASI and converted the square lattice into a rectangular lattice. In those works, the vertex interactions change with the length of the lattice constant, leading to tunable ground states in the rectangular ASIs as compared to the square ASI. In our work, the vertex interaction is tuned by changing the length of selected nanobars, while maintaining the Kagome lattice structure in all our samples.

# Supporting figures

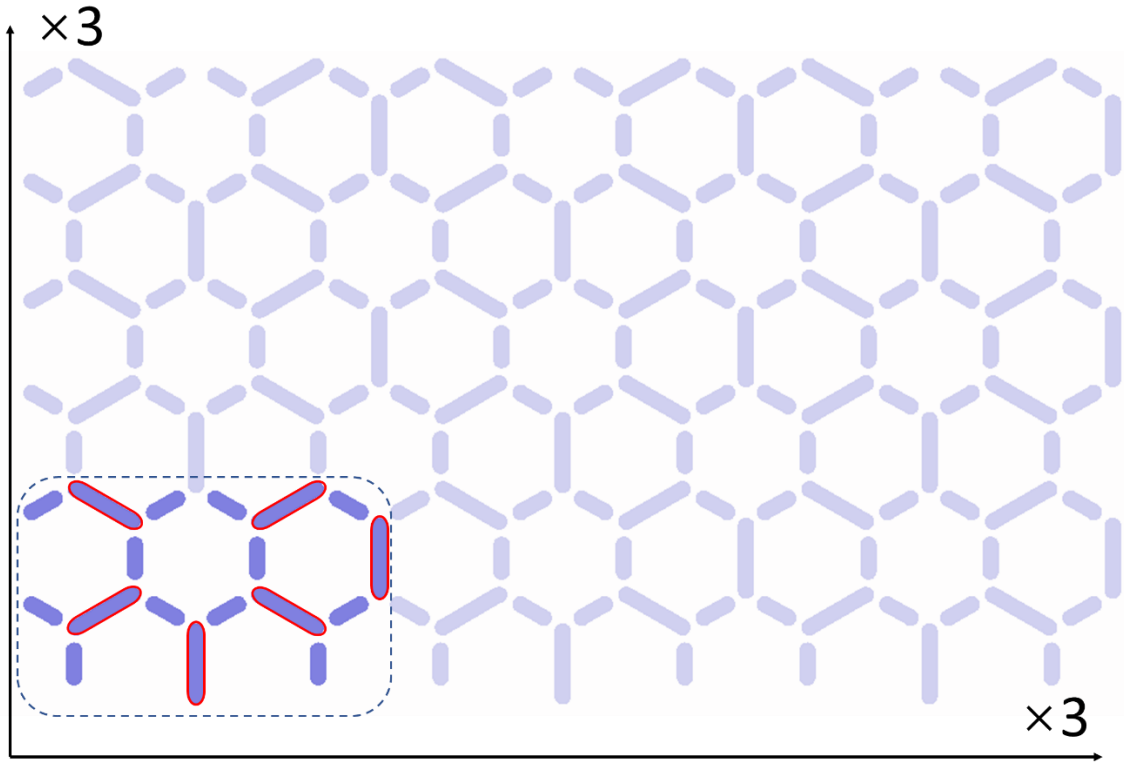

**Figure S1 | Sample design of Kagome ASI.** The dashed box in the bottom left is the basic repeating unit of the sample design. The $α$ magnets with adjustable length are highlighted with red boundaries.

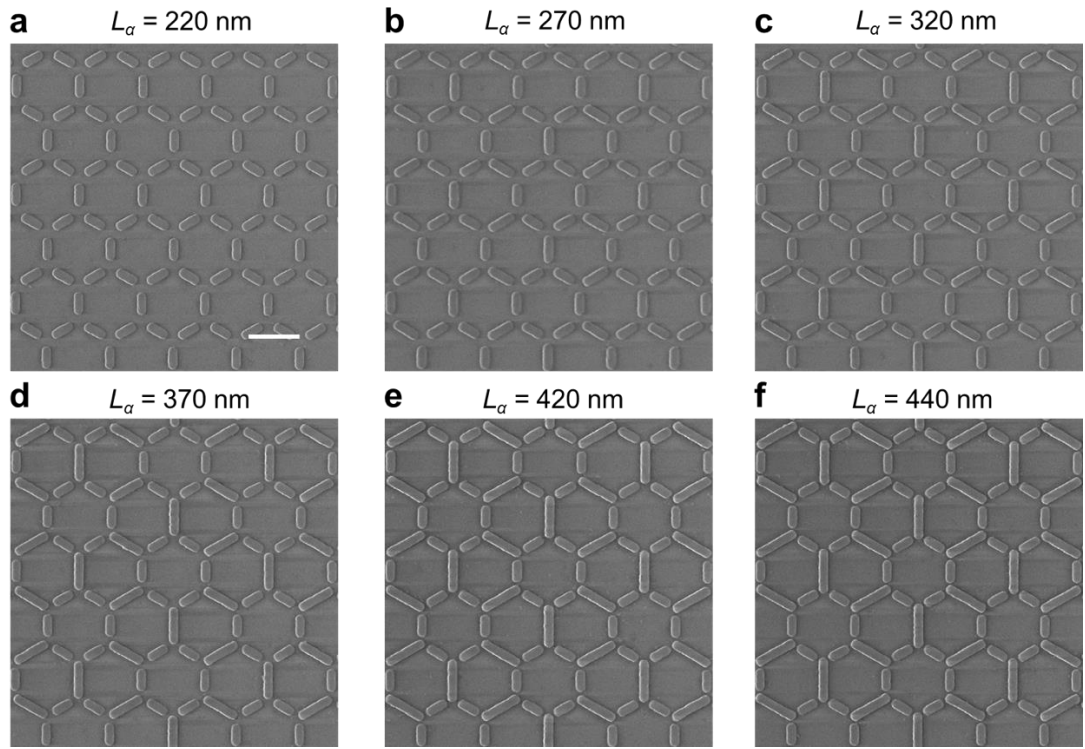

**Figure S2 | SEM images of six samples with various $L_a$ values.** Scale bar, 500 nm.

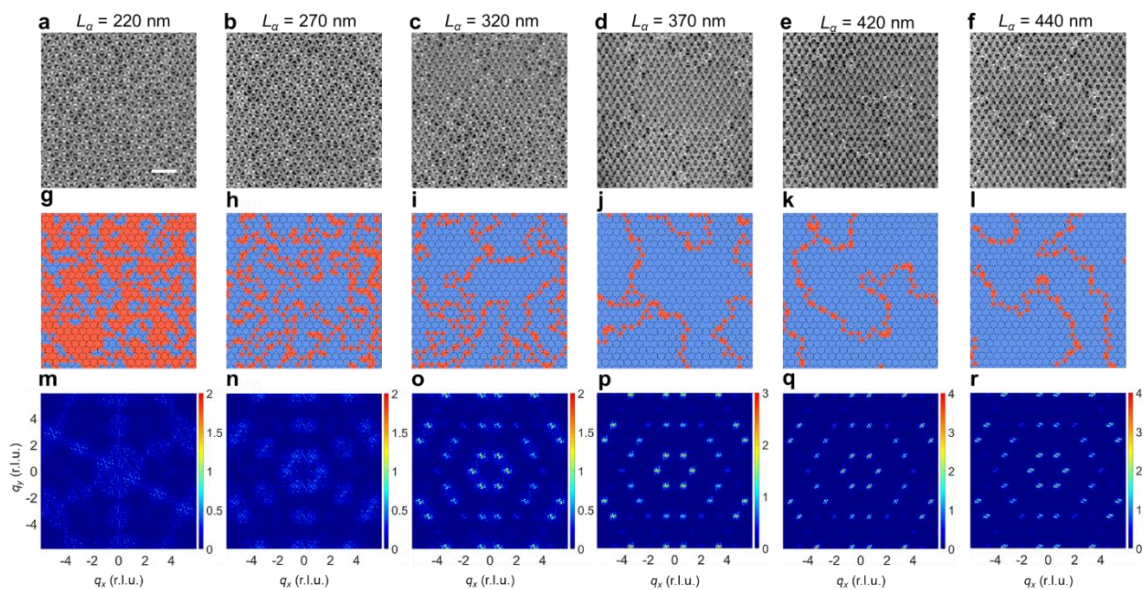

Figure S3 | **Experimental results of spin/moment configurations for all six measured samples. a-f**, MFM images. Scale bar, 2 μm. **g-h**, spin configurations and vertex distributions extracted from (a-f). Type K-I and K-II vertices are shown in blue and red, respectively. **m-r**, corresponding maps of magnetic structure factors calculated respectively from the spin configurations in (g-h).

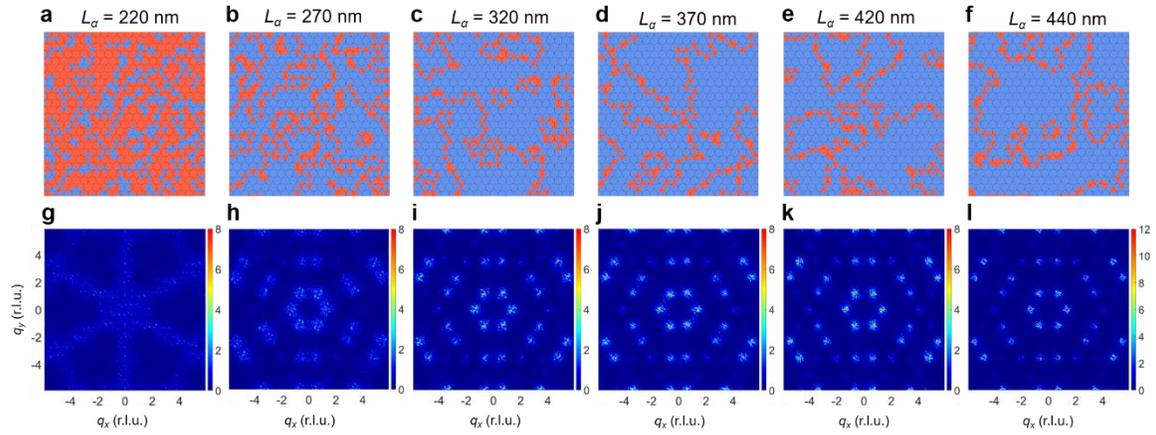

Figure S4 | **Monte Carlo simulation results of spin/moment configurations for all six samples. a-f**, spin configurations and vertex distributions obtained from Monte Carlo simulations. Type K-I and K-II vertices are shown in blue and red, respectively. **g-l**, corresponding maps of magnetic structure factors calculated respectively from the spin configurations in (a-f).

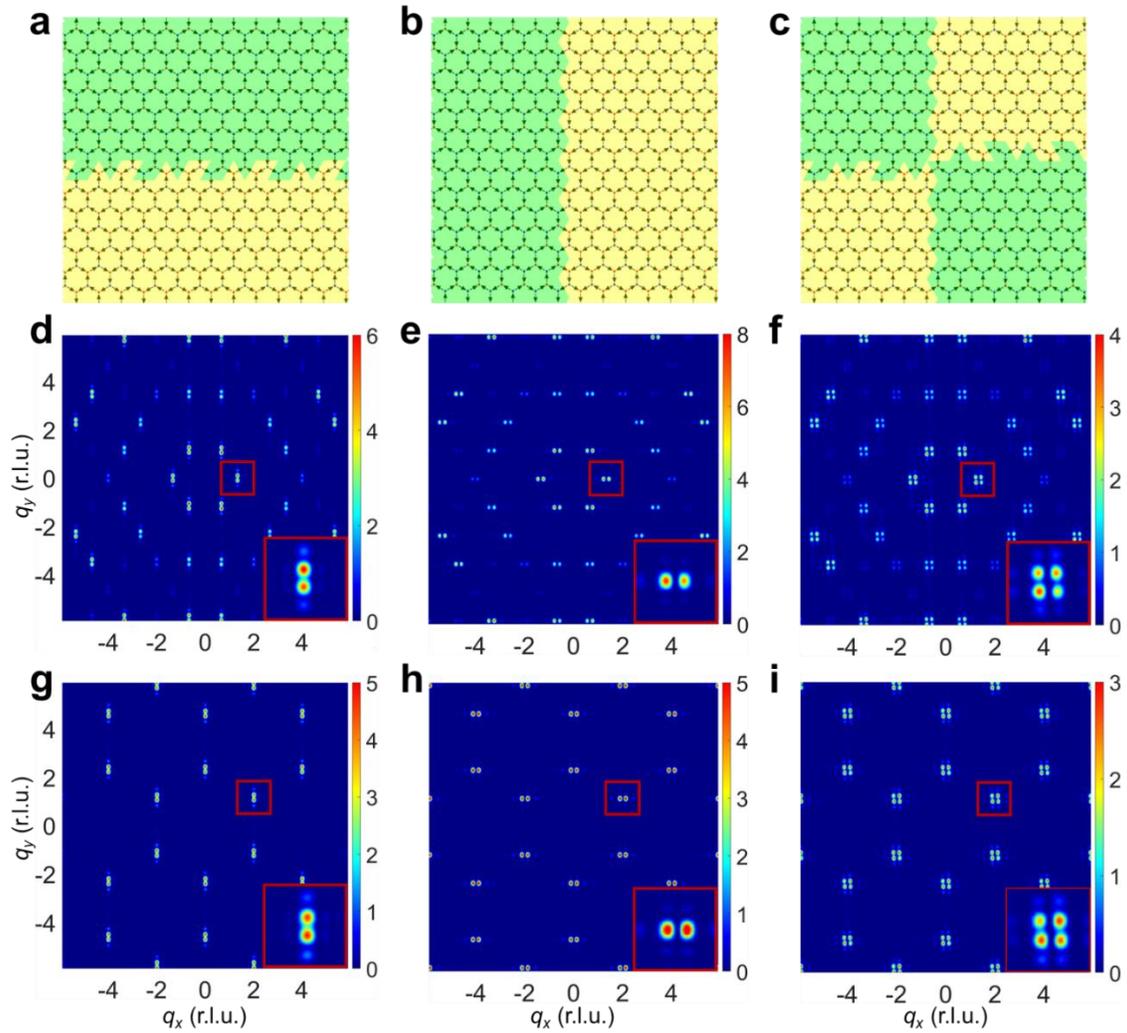

Figure S5 | **Domain induced Bragg peak splitting in magnetic structure factors. a-c**, three artificially designed spin/charge domain structures with horizontal (a), vertical (b) and cross (c) domain walls. Black arrows denote the spin configurations. Red and blue dots represent positive and negative magnetic charges. Two charge phases are encoded by green and yellow colors, respectively. **d-f**, maps of spin structure factors calculated respectively from the spin configurations in (a-c). **g-i**, maps of charge structure factors calculated respectively from the charge configurations in (a-c). Insets in (d-i) are expanded views of the splitting of Bragg peaks.

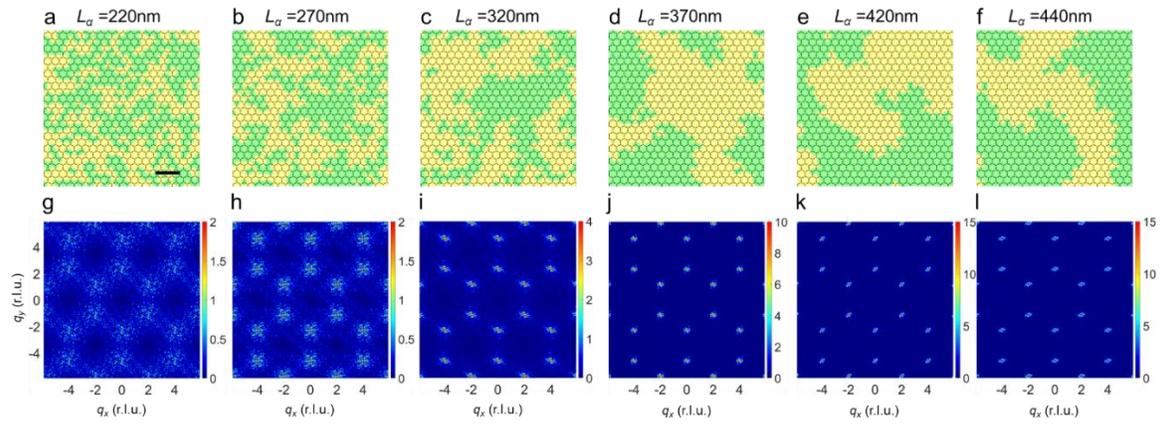

Figure S6 | **Experimental results of magnetic charge configurations for all six measured samples. a-f**, magnetic charge distributions corresponding to MFM images in Figs. S3a-S3f, respectively. Green and yellow denote two phases of magnetic charge ordering. **g-l**, maps of magnetic charge structure factors associated with (a-f).

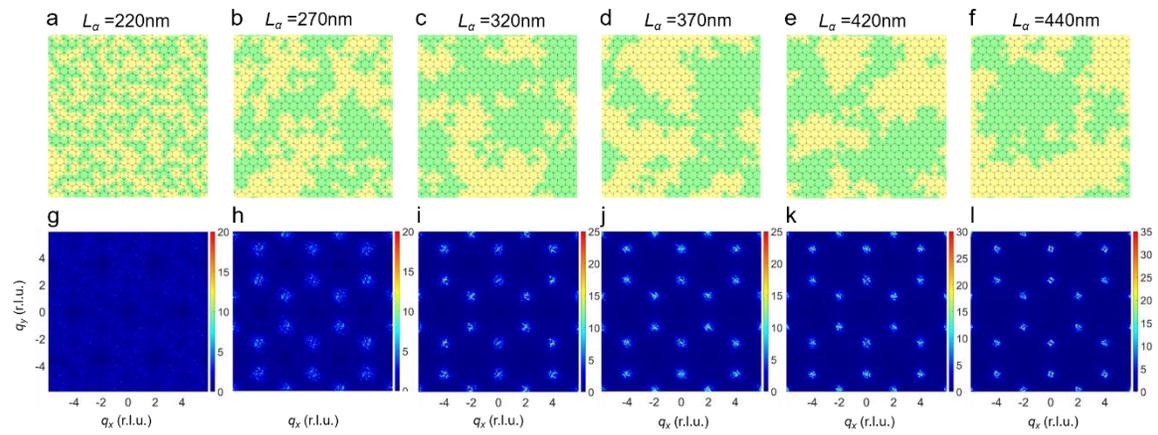

Figure S7 | **Monte Carlo simulation results of magnetic charge. a-d**, magnetic charge distributions. Green and yellow denote two phases of magnetic charge ordering. **g-l**, maps of magnetic charge structure factors associated with (a-f).

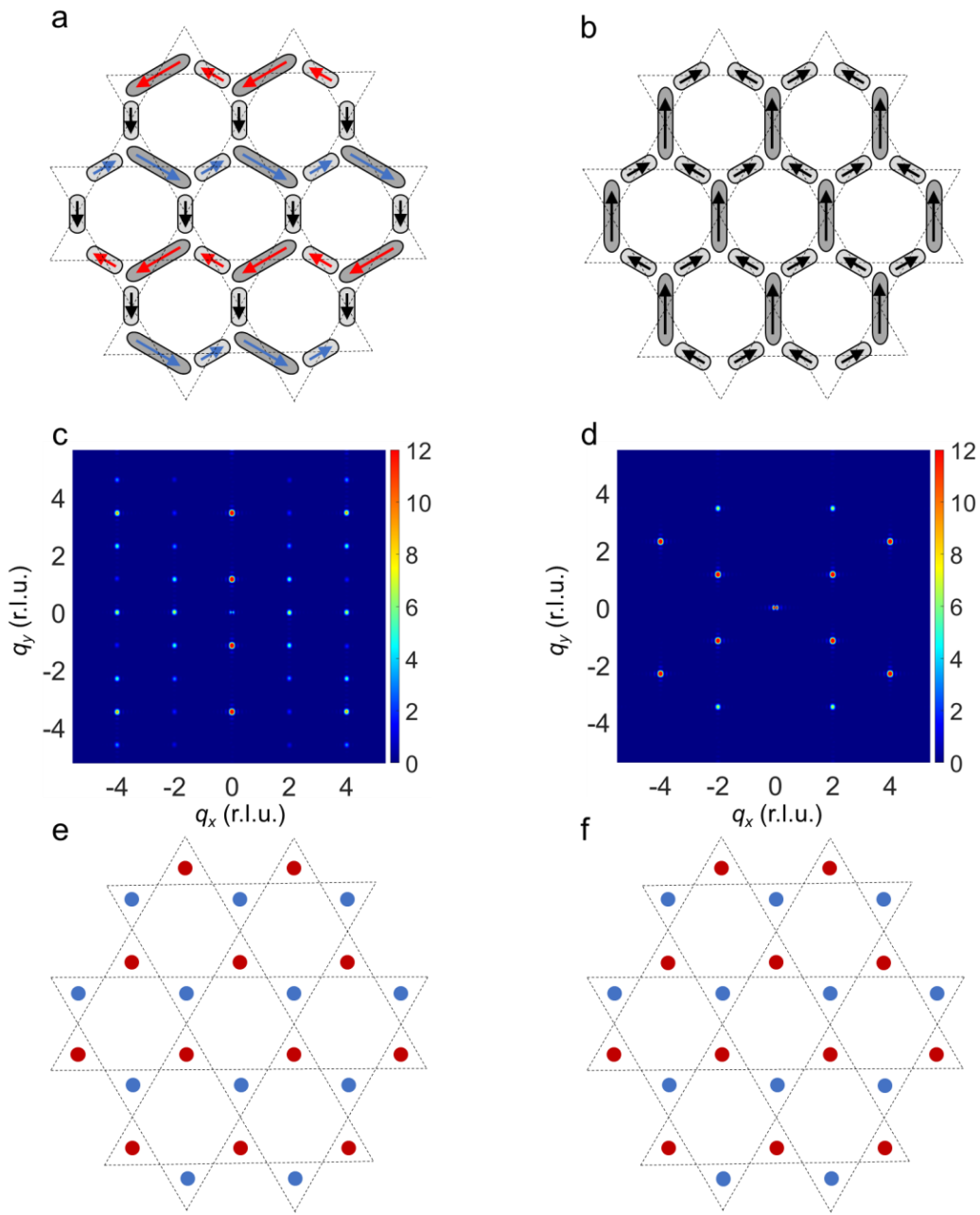

Figure S8 | **Examples of novel *ground states* in the modified Kagome ASIs. a**, an antiferromagnetic stripe-like ground state with red spins and blue spins point to right and left, respectively. **b**, a polarized ground state with the spins overall pointing up. *Although this spin configuration can also be realized as a metastable excited state in a standard Kagome ASI using external polarizing magnetic field, the spin configuration shown in this design is in the ground state.* In both (a) and (b), all the vertices satisfy the Type K-I configurations. **c** and **d**, the corresponding spin structure factor maps of (a) and (b), respectively. **e** and **f** show the magnetic charge ordering for (a) and (b).

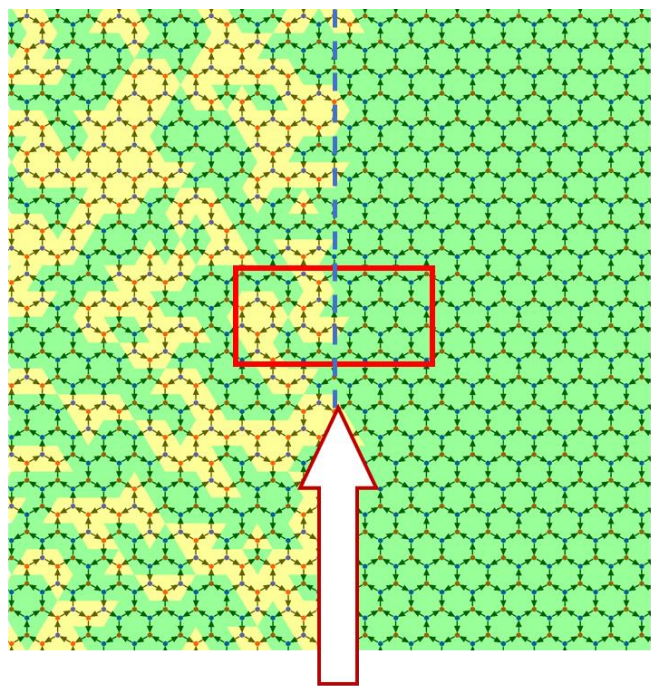

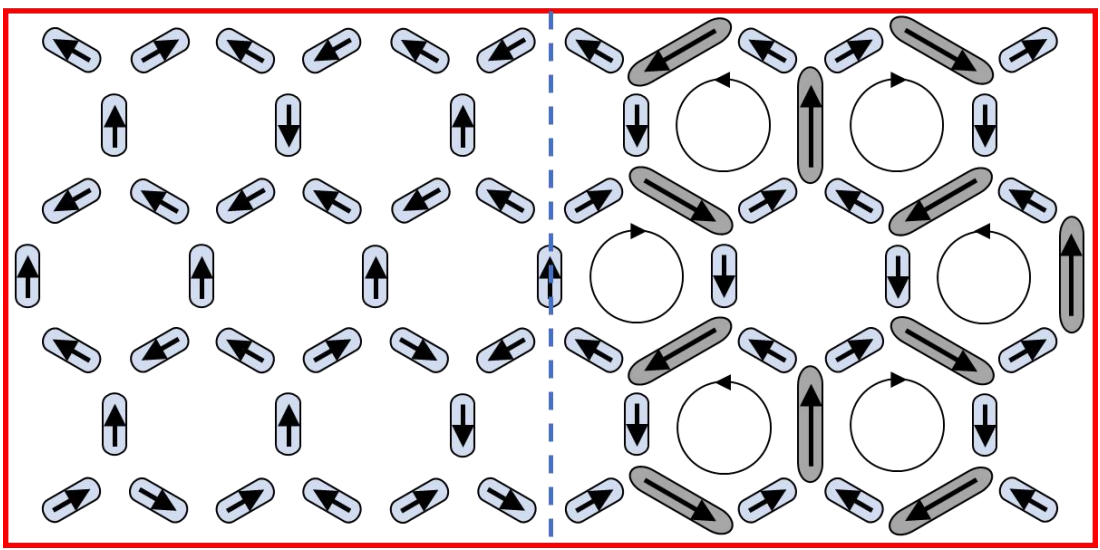

Figure S9 | **Kagome ASI with composite ground state.** The top figure shows spin and charge configurations of the ground state with spin liquid and spin crystal on left and right sides in the same Kagome lattice. Green and yellow colors denote two phases of magnetic charge. The bottom figure is an expanded view of the arrangement of nanomagnets within the red frame of the top figure.